\documentclass[sn-mathphys,Numbered]{sn-jnl}% Math and Physical Sciences Reference Style
\usepackage{graphicx}%
\usepackage{multirow}%
\usepackage{amsmath,amssymb,amsfonts}%
\usepackage{amsthm}%
\usepackage{mathrsfs}%
\usepackage[title]{appendix}%
\usepackage{xcolor}%
\usepackage{textcomp}%
\usepackage{manyfoot}%
\usepackage{booktabs}%
\usepackage{algorithm}%
\usepackage{algorithmicx}%
\usepackage{algpseudocode}%
\usepackage{listings}%
\usepackage{lineno}
\usepackage{upgreek}
\usepackage{amssymb}

\theoremstyle{thmstyleone}%
\theoremstyle{thmstyletwo}%

\theoremstyle{thmstylethree}%

\begin{document}

\title[Article Title]{On-chip Lithium Niobate Heterogeneous Photonic Crystal Nanocavity Laser}

%%=============================================================%%
%% Prefix	-> \pfx{Dr}
%% GivenName	-> \fnm{Joergen W.}
%% Particle	-> \spfx{van der} -> surname prefix
%% FamilyName	-> \sur{Ploeg}
%% Suffix	-> \sfx{IV}
%% NatureName	-> \tanm{Poet Laureate} -> Title after name
%% Degrees	-> \dgr{MSc, PhD}
%% \author*[1,2]{\pfx{Dr} \fnm{Joergen W.} \spfx{van der} \sur{Ploeg} \sfx{IV} \tanm{Poet Laureate} 
%%                 \dgr{MSc, PhD}}\email{iauthor@gmail.com}
%%=============================================================%%

\author[1,2]{\fnm{Xiangmin} \sur{Liu}}
\author[1,2]{\fnm{Rui} \sur{Ge}}

\author[1,2]{\fnm{Chengyu} \sur{Chen}}

\author[1,2]{\fnm{Jiangwei} \sur{Wu}}

\author*[1,2,3]{\fnm{Yuping} \sur{Chen}}\email{ypchen@sjtu.edu.cn}

\author[1,2,4,5]{\fnm{Xianfeng} \sur{Chen}}

\affil*[1]{\orgdiv{State Key Laboratory of Advanced Optical Communication Systems and Networks}, \orgname{School of Physics and Astronomy}, \orgaddress{\street{Shanghai Jiao Tong University}, \city{Shanghai}, \postcode{200240}, \country{China}}}

\affil[2]{\orgdiv{Shanghai Research Center for Quantum Sciences}, \orgaddress{\city{Shanghai}, \postcode{201315}, \country{China}}}

\affil[3]{\orgdiv{School of Physics}, \orgname{Ningxia University}, \orgaddress{\city{Yinchuan}, \postcode{750021}, \country{China}}}

\affil[4]{\orgdiv{Collaborative Innovation Center of Light Manipulations and Applications}, \orgname{Shandong Normal University}, \orgaddress{\city{Jinan}, \postcode{250358}, \country{China}}}

\affil[5]{\orgdiv{Shanghai Research Center for Quantum Sciences}, \orgaddress{\city{Shanghai}, \postcode{201315}, \country{China}}}

%%==================================%%
%% sample for unstructured abstract %%
%%==================================%%

\abstract{Thin film lithium niobate (TFLN) has become an platform for modern integrated circuits due to its excellent optical properties. With the development of rare earth ion doped TFLN, important breakthroughs of on-chip microlasers has emerged and show significant application for optical communication, computing and quantum photonics. However, challenges still remain in developing compact lasers with small mode volumes and low threshold on rare earth ion doped TFLN for highly efficient dense integration. In this letter, we fabricated a heterogeneous photonic crystal (PhC) nanobeam cavity on erbium-doped TFLN with a quality factor of $1.2\times 10^4$ and firstly demonstrated single-mode integrated PhC laser with submicron mode volume on TFLN platform. Laser at 1559.63 nm is achieved with the 974 nm single-mode pump. The effective mode volume is 1.44 $(\lambda/n)^3 (\sim 0.645 \upmu m^3)$ and the threshold power is 163 $\upmu$W. This lithium niobate photonic crystal nanocavity laser, as a compact telecommunication C-band on-chip light source, would benefit the progress of high-speed and low-cost optical community on TFLN integrated photonics.

Thin film lithium niobate (TFLN) has become an platform for modern integrated photonic circuits due to its excellent optical properties. With the development of rare earth ion doped TFLN, important breakthroughs of on-chip microlasers has emerged and show significant application for optical communication, computing and quantum photonics. However, challenges still remain in developing compact lasers with small mode volumes and low threshold on rare earth ion doped TFLN for highly efficient dense integration. In this letter, we fabricated a heterogeneous photonic crystal (PhC) nanobeam cavity on erbium-doped TFLN with a quality factor of $1.2\times 10^4$ and firstly demonstrated a single-mode integrated PhC laser with submicron mode volume on TFLN platform. Laser at 1559.63 nm is achieved with the 974 nm single-mode pump. 1.44 $(\lambda/n)^3 (\sim 0.645 \upmu m^3)$ and the threshold power is 163 $\upmu$W. Enhanced photorefractive effect in the PhC cavity is measured. This lithium niobate photonic crystal nanocavity laser, as a compact telecommunication C-band on-chip light source, would benefit the progress of high-speed and low-cost optical community on TFLN integrated photonics.  }


%%\pacs[JEL Classification]{D8, H51}

%%\pacs[MSC Classification]{35A01, 65L10, 65L12, 65L20, 65L70}

\maketitle

\section{Introduction}\label{sec1}

Modern high-speed and scalable optical communication and information processing urgently require the development of complete photonic integrated circuits (PICs) to break the bottleneck limited by bandwidth and power density. Thin film lithium niobate (TFLN) with attractive photoelectric and nonlinear properties (e.g.,wide transparent window from visible to mid-infrared, large electro-optical coefficient and second-order nonlinear coefficient) has become one of the most versatile material platform for integrated photonics\cite {ref1,boes2018status}. Within the past few years, various miniaturized and low-cost photonic devices with impressive performance have been demonstrated on TFLN, such as electro-optical and acousto-optic modulators, frequency converter and combs. \cite {wang2018integrated, he2019high,  lin2019broadband, sarabalis2020acousto, yu2021gigahertz, zhang2019broadband, wang2019monolithic}. In addition to these functional devices, on-chip light source, as an essential part of complete PICs, needs to be urgently achieved on TFLN whereas lithium niobate has indirect bandgap. Recently, the development of commercial rare earth ion doped TFLN (REI:LNOI) (e.g., erbium-doped, ytterbium-doped or erbium–ytterbium-co-doped) provides a solution to this conundrum and enables TFLN has gain properties. Plenty of researches based on REI:TFLN have been reported, for instance, microcavity laser, waveguide amplifier, soliton combs and  single photon sources\cite {lin2022electro, liu2021chip, wang2021chip, yin2021electro, luo2021chip, zhou2021chip1,  liu2021tunable, zhou2021chip2, yang1550, wang2023integrated, yang2023controlling}.

The devolopment of compact and low-cost light-emitting source is a significant research topic for inter/intra-chip communication on TFLN platform. Futhermore, microcavity produces an enhancement of photon emission proportional to Q/V based on Purcell effect, where Q is the quality factor and V is the mode volume\cite{sauvan2013theory}. Presently most microlasers on REI:TFLN described above are based on whispering gallery mode (WGM) microcavities. Limited by the total internal reflection condition, the radius of high Q WGM cavity should be large enough otherwise the scattering loss will increase. The corresponding effective V of reported WGM microlasers on REI:TFLN are typically tens or hundreds of cubic microns. \cite {lin2022electro, liu2021chip, wang2021chip, yin2021electro, luo2021chip, zhou2021chip1, liu2021tunable}. Thus the development of a novel type of on-chip light source with ultra-small mode volume and low energy consumption is compulsory for dense integration on TFLN.

Photonic crystal (PhC) nanocavities, which confine photons by the localizing defects in one-dimensional or two dimensional photonic lattices, have become a feasible solution for ultra-small on-chip light sources.\cite {istrate2006photonic}. The effective mode volume of PhC nanocavity is $\sim(\lambda/n)^3$, where $\lambda$ is the resonant wavelength and n is the effective refractive index. Various PhC lasers have been achieved in multiple platform (e.g., {\uppercase\expandafter{\romannumeral3}}-{\uppercase\expandafter{\romannumeral5}} semiconductors, perovskite and silicon-based hybrid laser) and demonstrate improved performance of device footprint and lasing threshold \cite {zhang2010photonic, gong2010nanobeam, trivino2015continuous, lee2017printed, li2017room, he2020cmos, fong2019silicon, wilhelm2017broadband, zhang2017advances}. Recently plenty of structural designs for improving Q/V of PhC nanocavity have been proposed and show potential application for low threshold lasers, single photon sources and cavity quantum electrodynamics, such as topological bulk laser, edge-state laser and corner-state laser, bichromatic PhC laser, and Magic-angle laser, etc. \cite {ota2018topological, zhang2020low1, shao2020high, simbula2017realization, mao2021magic, luan2023reconfigurable}. On TFLN platform, the fabrication of PhC is difficult because lithium niobate is hard to etch. With the improvement of process in recent years, high Q PhC nanocavities on TFLN have been reported and demonstrate impressive performance on nonlinear frequency conversion, high harmonic optomechanical oscillations and ultra-small electro-optic modulator\cite {liang2017high, jiang2018nonlinear, jiang2020high, li2020lithium}. However, PhC nanolaser based on REI: TFLN has not been achieved so far.

In this letter, we design and fabricate a PhC nanobeam cavity with V less than 1 $\upmu\rm m^3$ which supports resonance at C-band on erbium-doped TFLN (Er: TFLN). The single mode lithium niobate photonic crystal laser (LNPCL) with a center wavelength at 1559.63 nm is obtained with a 974 nm single-mode pump. The threshold power of LNPCL is 163 $\upmu$W and the simulated mode volume is 1.44 $(\lambda/n)^3$. The maximal slope efficiency is $4.15\times 10^{-6}\%$ and the maximal output power is 72 pW with pump power of 6.28 mW. The resonant wavelength of the PhC cavity shows a thermal response of 76 pm/$\rm^o$C. In addition, the resonance of LNPCL is changed by -1.38 nm/mW when regulating the pump power. This compact LNPCL can provide significant functionalities for densely integrated photonics on TFLN combined with electro-optic effect in the future.

\section{Design and Fabrication of Heterogeneous Photonic Crystal Cavity}\label{sec2}

\begin{figure}[h]%
\centering
\includegraphics[width=\textwidth]{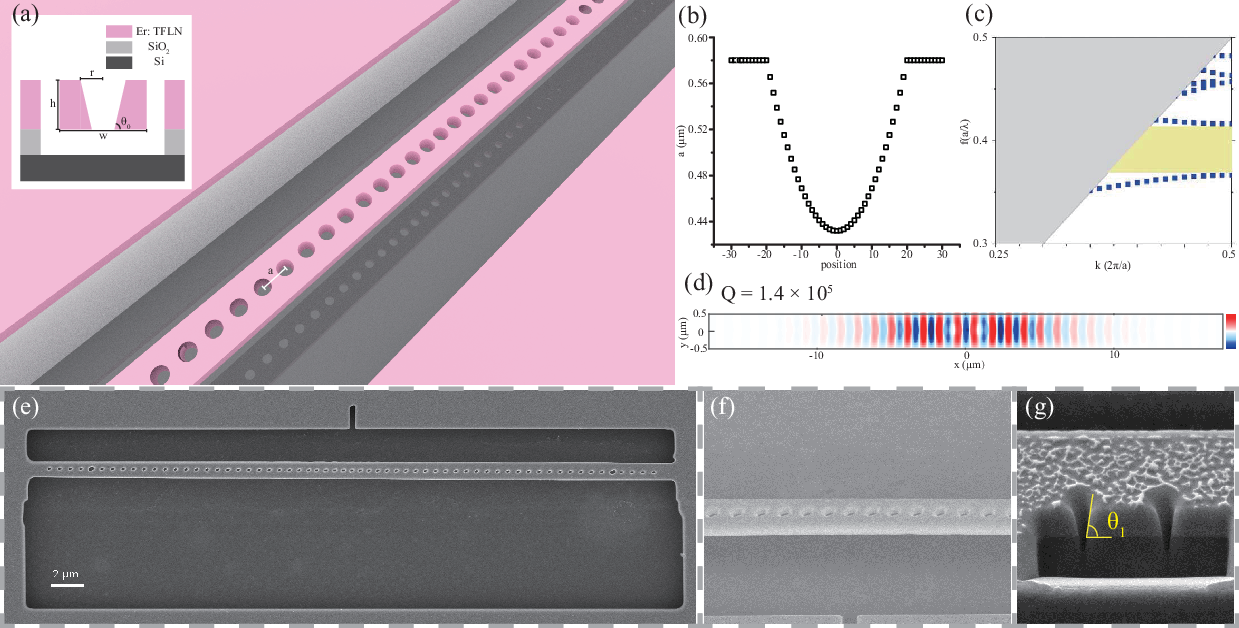}
\caption{(a) Schematic diagram of the LNPCL. Inset: The cross section of the nanobeam. a: lattice constant. r: radius of air holes. h: thickness of Er: LN. w: width of nanobeam. $\theta_0$: sidewall angle of air hole. (b) Relationship between the lattice constant and the position of holes. (c) Band diagram of the LN PhC nanobeam calculated by FDTD method. The blue square shows the dielectric and air bands and the yellow region shows the band gap. The gray region represents the light cone. (d) The simulated $\rm H_z$ component of  field profiles of the resonant optical mode with Q = $1.1 \times 10^5$. (e-f)  The scanning electron microscope (SEM) image of the PhC nanobeam cavity on the erbium-doped TFLN. (g) The cross section of the PhC nanobeam before removing Au and HF etching, illustrating the slant angle of the air hole is $\theta_1 = 80.9^{\circ}$. }\label{fig1}
\end{figure}

Fig. 1(a) illustrates the schematic of our PhC nanobeam cavity on Er: TFLN. Our device is cladded with air and consists of lattice constant taper of ten periods arranged symmetrically about the center of the cavity. To produce a defect cavity, the lattice constants shown in Fig. 1(b) are parabolic with the position coordinates and vary from 580 nm to 432 nm. According to the cross section of nanobeam in the inset of Fig. 1(a), the designed top radius of air holes is $r=168$ nm. The width and the layer thickness of the nanobeam is seperately $w=1$ $\upmu$m and $h=250$ nm. Based on the ion etching process for fabricating TFLN devices, we consider the sidewall angle of the air hole is $\theta_0 = 80^{\circ}$.

As shown in Fig.1(c), we simulate the band structure of the PhC nanobeam cavity via meep, which is an open-source software package of the three dimensional finite-difference time-domain (FDTD) method. \cite{oskooi2010meep}. Blue dots in Fig.1(d) displays the dispersion properties of the PhC nanobeam, while the gray region is light cone and the yellow region indicates the band gap. By structural design, the band gap with a width of 21.6 THz locates in the telecom band and the transverse-electric (TE) guide mode is well confined below the light line. With the detailed simulation, this PhC cavity support the $\rm TE_{01}$ mode at a wavelength of 1532.33 nm and the corresponding calculated Q is $1.4 \times 10^5$. The distribution of $\rm H_z$ component of the $\rm TE_{01}$ mode is demonstrated in Fig.1 (d). The calculated mode volume V of this cavity mode is 1.44 $(\lambda/n)^3 \approx 0.645 \upmu \rm m^3$, and the physical volume of the whole nanobeam is $45\times1\times0.25 \upmu m^3(\sim 25.1 (\lambda/n)^3)$. Pump light around 1480 nm locates in the photonic bandgap although it has a higher slope efficiency. Thus the 980 nm-pump locating at guide mode is used to generate laser. Based on the energy level of erbium ion, the up-conversion signal light at 1.53 $\upmu$m resonates in the PhC cavity and finally emits with pump light around 980 nm.

The PhC nanobeam is fabricated on an 1$\%$ mol erbium-doped Z-cut TFLN wafer. Fig. 1(e) shows the SEM image of the sample and Fig. 1(f) is the locally enlarged image of the nanobeam. The pattern of PhC nanobeam cavity was fabricated by focused ion beam (FIB) etching and the air-clad nanobeam is prepared after removing silica layer. Fig. 1(g) shows the cross section of the nanobeam cut by FIB before hydrofluoric acid (HF) etching and the measured sidewall angle of the air hole is $\theta = 80.9^{\circ}$, which is close to our simulation settings. Detailed fabrication process are described in the method.

\section{Experimental setup}\label{sec3}
\begin{figure}[h]%
\centering
\includegraphics[width=\textwidth]{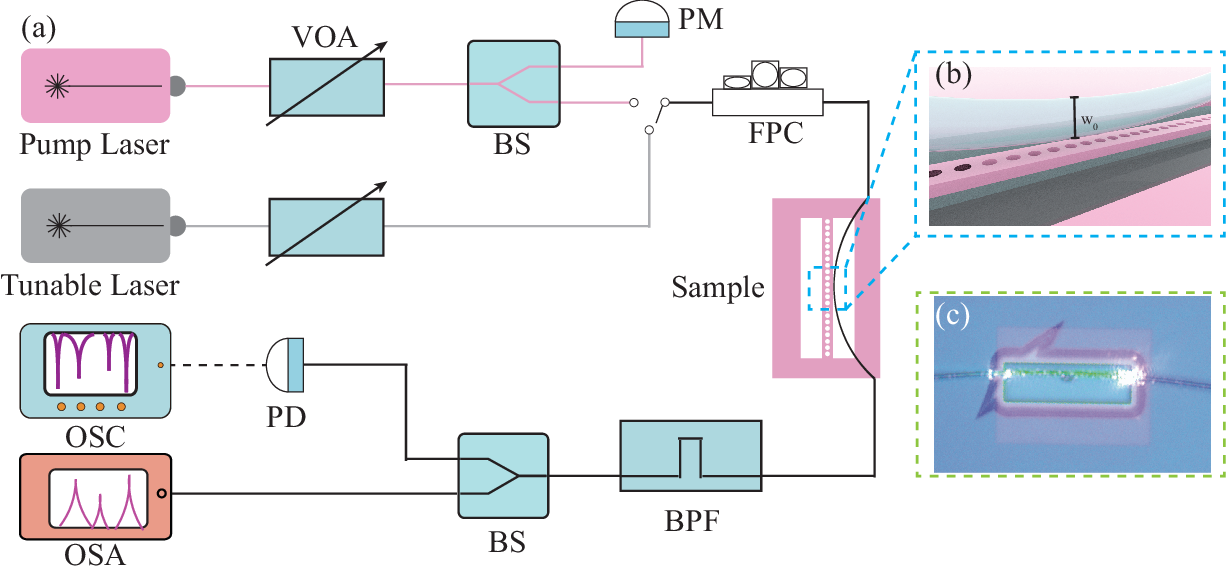}
\caption{(a) Schematic diagram of the experimental setup. Inset: the optical microscope images of the nanobeam with pump. VOA, variable optical attenuator; BS: beam splitter;  FPC, fibre-polarization controller; PM, power meter; PD, photodetector;BPF, bandpass filter; OSA, optical spectrum analyzer; OSC, oscilloscope. (b) Enlarged image of coupling between taper fiber and nanobeam. (c) Optical microscopic image of LNPCL under 974 nm pump.}\label{fig2}
\end{figure}

The schematic diagram of experimental setup is shown in Fig. 2(a). A 974 nm single-mode LD light source (Golight Co., Ltd) is used as the pump source to generate laser emission in the PhC cavity on the Er: TFLN. The pump light after attenuation propagates through the PC and  then couples into the nanobeam via the taper fiber with a waist of 1.8 $\upmu$m. A 99:1 BS and power meter are used to measure the pump power. Then signal light are collected by the optical spectrum analyzer to obtian laser spectra. To obtain the transmission spectrum of the PhC cavity, probe light generated by a tunable continuous-wave laser (New focus, TLB-6700) from 1520 nm to 1570 nm is attenuated by VOA to avoid thermo-optical effect and finally measured by PD and oscilloscope. Fig. 2(b) shows the coupling between taper fiber and nanobeam and the coupling condition is adjusted by moving the position of the taper fiber relative to the nanobeam. During the experiment, when pumped by the 974 nm laser, bright green up-conversion fluorescence due to the cooperation up-conversion (CUC) and excited-state absorption (ESA) of the pump light is observed in the nanobeam from the optical microscopic image shown in Fig. 2(c), which illustrates that the pump light is coupled into the nanobeam. 

\section{Results and discussion}\label{sec4}

\begin{figure}[h]%
\centering
\includegraphics[width=0.9\textwidth]{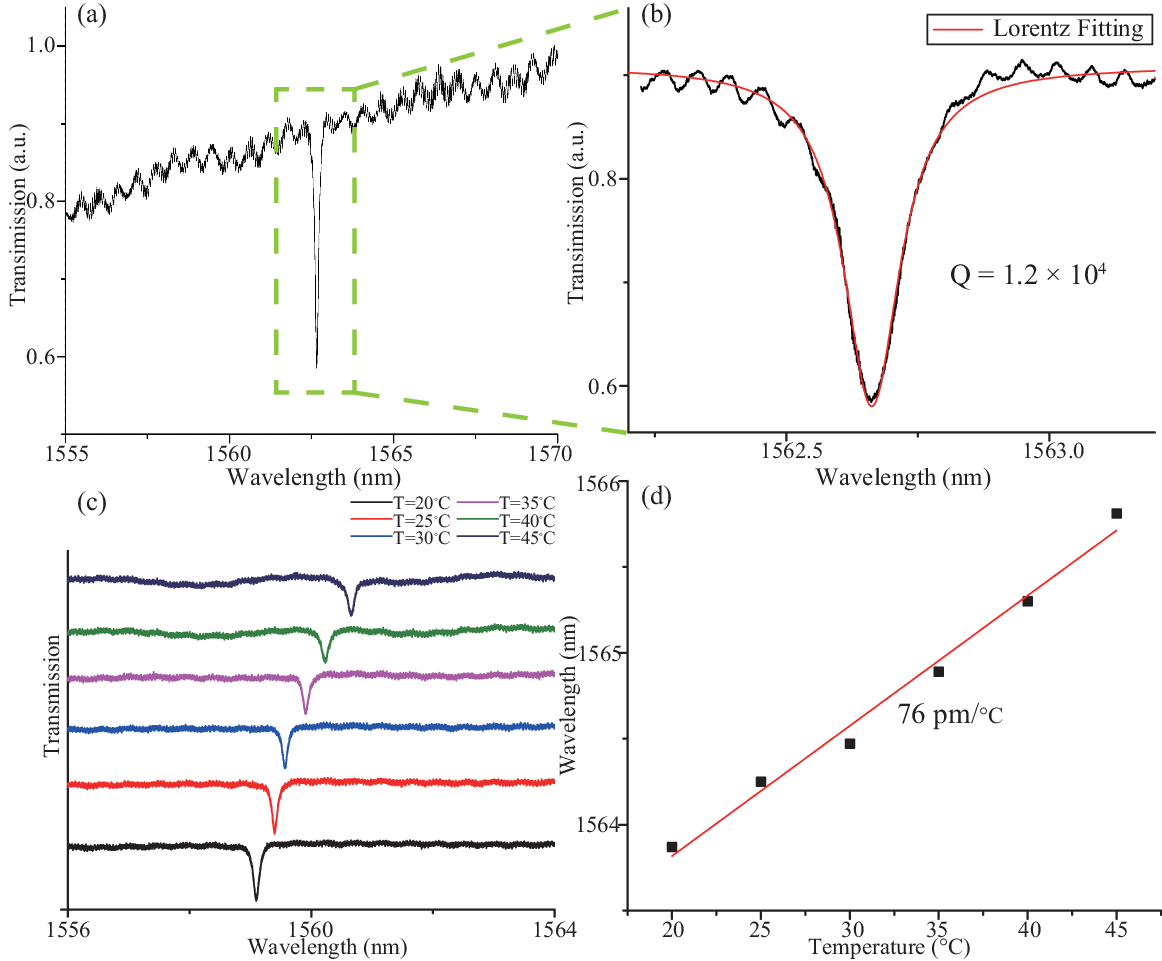}
\caption{(a)  The transmission spectrum of the PhC laser in erbium-doped TFLN from 1555 nm to 1570 nm. (b) The red curve shows the Lorentz fitting of the measured mode at 1562.67 nm in the green dashed frame in (a), indicating a Q factor of $1.2\times 10^4$. (c) The trasmission spectrum under different temperature. (d) The resonant wavelength versus different temperature. }\label{fig3}
\end{figure}

Fig. 3(a) demonstrates the measured transmission spectrum of the PhC nanobeam cavity from 1555 nm to 1570 nm at room temperature $T = 25 \rm^o$C. We found only one resonant cavity mode at the wavelength of 1562.67 nm in the C band and is located near the gain region of erbium ion around 1560 nm , as shown in the green dashed frame in Fig. 3(a). The corresponding Q factor is $1.2\times 10^4$ evaluated by Lorentz fitting, which shows the PhC nanobeam cavity on Er: TFLN support single-mode LNPCL emission around 1560 nm under pump. The transmission spectrum under different temperature is shown in Fig. 3(c). The PhC cavity with small V demonstrates a strong thermo-optical effect. The resonant wavelength redshifts from 1563.87 nm to 1565.81 nm when temperature increases from  $20 \rm^o$C to $45 \rm^o$C. The PhC cavity demonstrates a thermal response of 76 pm/$\rm^o$C. The taper fiber is adjusted to the critical coupling position at T=$20 \rm^o$C. With increased temperature, the coupling depth is decreased due to the change of the distance between taper fiber and nanobeam caused by the thermal expansion of LN.

\begin{figure}[h]%
\centering
\includegraphics[width=0.9\textwidth]{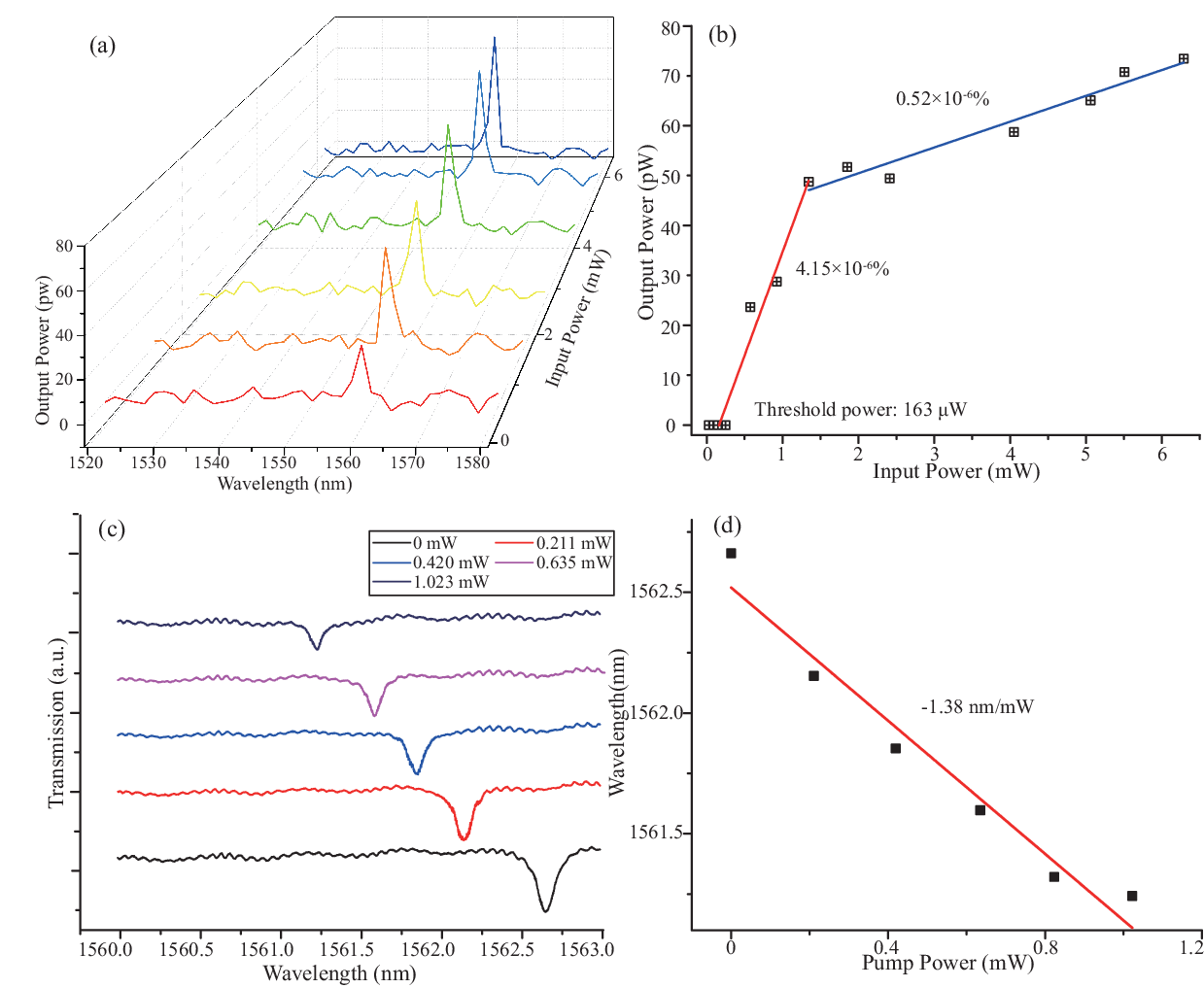}
\caption{(a) The spectra of PhC laser on erbium-doped TFLN under different wavelength and pump powers of 0.320mW, 1.345 mW, 2.413 mW, 4.048 mW, 5.504 mW and 6.285 mW. (b) The relationship between the laser power and the pump power. (c) The transmission spectrum around 1560 nm under different pump power. (d) The resonant wavelength of PhC cavity versus pump power.}\label{fig4}
\end{figure}

Fig. 4(a) illustrates the LNPCL emission spectra in the range of 1520 nm to 1570 nm under different power. The single-mode LNPCL is stable with the increase of pump power and the maximal output power is 72 pW with the pump power of 6.285 mW. Fig 4(b) demonstrates the output power versus the on-chip input power, which shows that the threshold power is 163 $\upmu$W. However, the laser wavelength deviates from the cavity mode measured in Fig. 3(a) and the PhC laser has different slope efficiency under different pump power. Compared with the thermo-optical effect, photorefractive effect plays a dominant role in wavelength shift. The center wavelength of LNPCL blueshifts from 1559.43 nm to 1556.56 nm when the pump power increases from 0 mW to 5.06 mW. Limited by the wavelength resolution of the spectrometer, the laser blueshift caused by the photonrefractive effect is difficult to measure accurately. Therefore, we demonstrate the response for photorefractive effect by measuring the transmission spectrum of PhC cavity at different pump power. 0.8 $\upmu$W singal light from 1556 nm to 1564 nm and pump light at 974 nm is simultaneously coupled into the PhC cavity and the transmission spectra under different pump power is shown in Fig. 4(d)). The resonant wavelegth of PhC cavity in Fig. 4(e) blueshifts with the input power and the slope is linear fitted to be -1.38 nm/mW. Additionally, the linewidth of resonant mode and coupling depth decrease with pump power because LNPCL emits. Due to the small size of the nanobeam, the strong photorefractive effect leads to the more obvious blue shift of the cavity mode than previous work of microring laser on Er: TFLN \cite{li2021single, liang2017high}. The central wavelength of LNPCL gradually deviates from the gain region of erbium ion around 1560 nm, thus suppressing the increase of laser power. Therefore we piecewise linear fit the laser power versus imput power at the same central wavelength as shown in Fig. 4(b). At the wavelength of 1558.09 nm, the LNPCL has a slope efficiency of $4.15 \times 10^{-6}\%$ and at the wavelength of 1556.56 nm, the slope efficiency is $0.52 \times 10^{-6}\%$, which shows the decreased slope efficiency with increased pump power due to blueshift of resonant wavelength. The strong photorefractive effect can be suppressed by magnesium-doped LN, and thus stable slope efficiency can be achieved. \cite {bryan1985magnesium}.

\begin{table}[h]
\caption{Parameters comparison of microlasers on Er:TFLN}\label{tab1}%
\begin{tabular}{@{}llllllll@{}}
\toprule
Ref. & Device & Footprint & Pump $\lambda$ & Q factor & mode volume & Threshold & Slope \\
    &     &      & (nm) &    & ($\upmu m^3$) & power (mW) & efficiency \\
\midrule

\cite{wang2021chip} & Microdisk & $d=200\upmu$m & 976 &  $1.02\times 10^6$ & N/A & $\sim0.25$ & $1.92 \times 10^{-4}$ \\
\cite{yin2021electro} & Racetrack ring & $400\upmu$m $\times 900\upmu$m & 980 &  $1.54\times 10^5$ & N/A &$\sim3.5 $ & $4.38 \times 10^{-5}$ \\
\cite{liang2022monolithic} & Microring & $750\upmu$m $\times 750\upmu$m & 976 &  $3.2\times 10^5$ & N/A &$\sim24.5 $ & $8.33 \times 10^{-6}$ \\
\cite{li2021single} & Microring & $210\upmu$m & 1484 &  $2.13\times 10^5$ & N/A &$\sim14.5 $ & $1.20 \times 10^{-4}$ \\
\cite{gao2021chip} & Coupled disk & $d_1, d_2=23.1, 29.8\upmu$m & 977.7 &  $1.6\times 10^6$ & N/A & $\sim0.2$ & $7.0 \times 10^{-5}$ \\
\cite{yu2023chip} & FP cavity & $6.5$mm $\times 1.5$mm & 980 &  $1.6\times 10^5$ & N/A &$\sim6$  & $0.18$ \\

\cite{liu2021chip}$^\ast$ & Microdisk & $d=150\upmu$m & 974 &  $1.05\times 10^5$ &$\sim170 $&$\sim2.99$  & $4.117 \times 10^{-6}$ \\
\cite{liu2021tunable}$^\ast$ & Coupled ring & $d_1, d_2=300, 330\upmu$m & 974 &  $1.7\times 10^5$ & $\sim488 $ &$\sim1.31$  & $4.41 \times 10^{-5}$ \\
Our work$^\ast$ & PhC & $45\upmu$m $\times 1\upmu$m & 974 &  $1.2\times 10^4$ &$\sim0.645$&$\sim0.163$ & $4.15 \times 10^{-6}$ \\

\botrule
\end{tabular}
\footnotetext[\ast]{Our previous work.}
\end{table}

Table 1 exhibits the comparison of parameters and properties of reported microlasers on Er:TFLN. Our LNPCL demonstrates the minimum footprint, which benefits more dense integration. In contrast to our previus work, the effective V of LNPCL significantly decreases and large Purcell factor Q/V leads to a lower laser threshold. However, compared with other microcavity laser on Er: TFLN, the LNPCL has lower slope efficiency. One the one hand, LNPCL has a smaller Q and the tiny footprint leads to much smaller gain region than WGM cavity. Therefore the fabrication process needs to be improved to obtain higher Q to improve the slope efficiency. On the other hand, WGM microcavity supports the resonance of pump light and laser simultaneously to enhance the intracavity pump power while 974 nm-pump locates in the guidemode in LNPCL. Lower intracavity pump power of LNPCL leads to increased threshold power and decreased slpoe efficiency. The structure of PhC can be optimized to support pump light around 1480 nm and laser resonant simultaneously with high Q mode in the PhC nanocavity to reduce energy consumption and promote laser efficiency.

\begin{figure}[h]%
\centering
\includegraphics[width=1\textwidth]{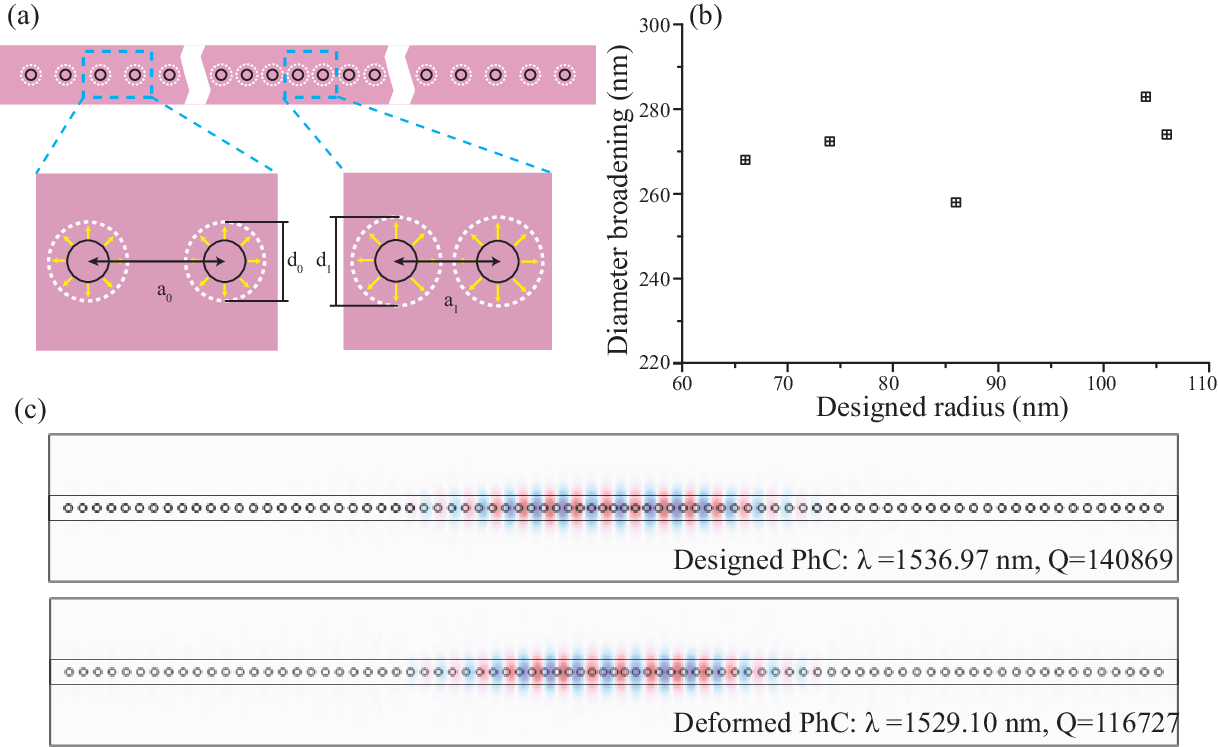}
\caption{(a) Schematic diagram of line broadening effect and promixity effect. The solid black circle is the designed air hole, and the dashed white circle is the air hole with increased radius after FIB etching when considering line broadening effect. The yellow arrow shows the increased radii. Lattice constants: $a_0=580 nm, a_1=432 nm$. Air hole diameter: $d_0=336 nm, d_1= 356 nm$. The difference in $d_1, d_2$ at different locations shows the effect of promixity effect.(b)Radius broadening under different designed radii. (c)  The simulated $\rm E_y$ component of field profiles of designed PhC and deformed PhC with promixity effect.
}\label{fig5}
\end{figure}

Various manufacturing imperfection produced by FIB will affect the resonant wavelength and Q of PhC cavity, such as line broadening effect and promixity effect shown in Fig. 4(a). Air hole radii after FIB milling will broaden due to line broadening effect caused by backscattering of ions\cite{lacour2005nanostructuring}. As shown in Fig. 4(b), we measure the linear broadening effect with same lattice constant and found the radius broadening at different designed radii from 64 nm to 106 nm is $\sim$270 nm, which has been considered in above design and fabrication process. In addition, line broadening effect is also affected by promixity effect. Area between air holes in close proximity is repeatedly exposed during FIB, resulting in an increased air hole radius with decreased lattice constant. We measured that the diameter of air hole with a lattice constant $a_0$=580 nm at mirror region is $d_0=$336nm and diameter air hole with $a_1$=432 nm at center is $d_1$=356nm. The promixity effect leads to the deformation of air holes and the diameter is increased by 20 nm at center. We simulated the $\rm E_y$ component of field profiles of designed PhC and deformed PhC. The deformation of the air hole results in the decreased Q value and the blue shift of resonant wavelength to 7.87nm. However, the resonant wavelength of deformed PhC is still located in the gain region of $\rm Er^{3+}$. The redshift of the measured resonant wavelength (1562.67 nm) relative to the design wavelength (1536.97 nm) also needs to consider other manufacturing imperfection, such as deviation of width of nanobeam and sidewall roughness, which are analyzed in the supplementary materials.

\section{Conclusion}\label{sec5}

In conclusion, we demonstrate the on-chip single-mode LNPCL on Er: TFLN with submicron mode volume for the first time. A PhC nanobeam cavity with the V of 1.44 $(\lambda/n)^3$ and a Q factor of $1.2\times 10^4$ is fabricated on Er: TFLN by FIB. The stable single-mode LNPCL is emitted at 1559.63 nm with a threshold power of 1.56 mW when pumped with 974 nm-single-mode laser. The maximal slope efficiency is $4.15 \times 10^{-6}\%$. The resonant wavelength of PhC cavity redshifts with a slope of 76 pm/$\rm^o$C with increased temperature. In addition, the blue shift of central wavelength and the change of slope efficiency of LNPCL due to photorefractive effect are also measured and analyzed with the increased pump power. Our PhC laser will be combined with the electro-optic effect of TFLN in the future research to more precisely control the laser. As a C-band compact light source with effective mode volume less than 1 $\upmu m^3$ on TFLN, the LNPCL shows great potential in high density lithium niobate photonic integrated circuits.

\section{Method}\label{sec6}
The PhC nanobeam is fabricated on an 1$\%$ mol erbium-doped Z-cut TFLN wafer. 1$\%$mol erbium ion is doped into LN during crystal growth process. The Er:TFLN consists of a 600-nm-thick erbium-doped thin film LN, a 2 $\upmu$m-silica buffer layer and a 400 $\upmu$m-silicon substrate is fabricated by the smart-cut method from a 3-inch wafer. The Er: TFLN wafer is cut into chips with the size of 0.8 mm$\times$0.8 mm. and The thickness of the Er: LN layer is reduced from 600 nm to 250 nm by chemical mechanical polishing (CMP) for 80s.

The detailed fabrication process of the PhC nanobeam cavity is shown in Fig.S2 in supplementary. (1) Paraffin on Er:TFLN after CMP was removed by alcohol. (2) The chip was wet cleaned with acetone, isopropyl alcohol and SPM solution. (3) Au was sputter onto the surface of Er: LN layer. (4) The structure of PhC nanobeam cavity was precisely engineered by SEM and a focused ion beam (FIB) dual beam system (ZEISS Auriga). The corresponding accelerated voltage of Ga$^+$ is 30 kV and the probe current of the electrostatic lenses is 120 pA. (5) After FIB milling, we immersed the sample into the aqua regia to remove Au. (6) We immersed the sample into a buffered oxide etching (BOE) solution for 35min to remove the silica layer under the PhC and obtained air-clad nanobeam.

\backmatter

\bmhead{Acknowledgments}

This work was supported by the National Natural Science Foundation of China (Grant. 12134009), Shanghai Municipal Science and Technology Major Project (2019SHZDZX01ZX06), and Shanghai  Jiao Tong University  Project (21X010200828).

\bibliography{sn-bibliography}% common bib file
%% if required, the content of .bbl file can be included here once bbl is generated
%%\input sn-article.bbl

\end{document}